\documentclass[pmlr]{jmlr}

\usepackage{longtable}

\usepackage{booktabs}
\usepackage[load-configurations=version-1]{siunitx} 

\theorembodyfont{\upshape}
\theoremheaderfont{\scshape}
\theorempostheader{:}
\theoremsep{\newline}

\jmlrvolume{1}
\jmlryear{2010}
\jmlrworkshop{Workshop Title}

\title{Against `softmaxing' culture}

  \author{\Name{Daniel Mwesigwa} \Email{dm663@cornell.edu}
   \addr Ithaca, NY}

\begin{document}

\maketitle

\begin{abstract}
AI is flattening culture. Evaluations of “culture” are showing the myriad ways in which large AI models are homogenizing language and culture, averaging out rich linguistic differences into generic expressions. I call this phenomenon ``softmaxing culture," and it is one of the fundamental challenges facing AI evaluations today. Efforts to improve and strengthen evaluations of culture are central to the project of cultural alignment in large AI systems. This position paper argues that machine learning (ML) and human-computer interaction (HCI) approaches to evaluation are limited. I propose two key conceptual shifts. First, instead of asking “what is culture?” at the start of system evaluations, I propose beginning with the question: “when is culture?” Second, while I acknowledge the philosophical claim that cultural universals exist, the challenge is not simply to describe them, but to situate them in relation to their particulars. Taken together, these conceptual shifts invite evaluation approaches that move beyond technical requirements toward perspectives that are more responsive to the complexities of culture.

\end{abstract}


\section{Introduction}
\label{sec:intro}

Large AI models, particularly large language models (LLMs), mediate how culture is represented, interpreted, and reproduced \citep{liu_culturally_2025,adilazuarda_towards_2024,zhou_culture_2025,bajohr_thinking_2025,farrell_large_2025}. These systems promise universal fluency and translation across linguistic and cultural boundaries, but in doing so, they often flatten the complexity of culture into generic forms. I refer to this phenomenon as “softmaxing culture.”\footnote{In ML, the \texttt{softmax function} transforms a set of raw output scores into a probability distribution, amplifying the most likely outputs while reducing the less probable. This probabilistic logic, when applied to \textit{culture}, has a clear effect: it privileges dominant, statistically frequent expressions while suppressing non-frequent ones. What is statistically probable is often confused with what is socially or culturally appropriate. Meanwhile, it is also important to note that the phenomenon I have described has nothing to do with `softmaxxing,' a beauty trend associated with enhancing one's appearance.} A rapidly expanding body of work is showing how LLMs are driving diverse users of English as a second language to Western linguistic norms and patterns \citep{bommasani_opportunities_2022,agarwal_ai_2025}; how LLMs propagate racial, geographic, and gender-based stereotypes \citep{cheng_marked_2023,decoupes_evaluation_2025}; and judgments by LLMs are prejudiced in scoring comparisons around diversity \citep{ye_justice_2024}. While NLP is grappling with the challenges of these large AI models, including the biases and stereotypes that these models propagate \citep{bender_dangers_2021}, the evaluation of these systems is limited \citep{wallach_position_2025}. This limitation stems from the fact that model outputs are statistical predictions of patterns learned from large corpora of training data. As such, system evaluations tend to measure narrowly defined aspects of language and culture, leaving out the more complex, relational aspects that constitute culture and language. 

Despite differences in method and practice between ML and HCI research,\footnote{It is true, ML research is re-inventing HCI’s core problematic – user research. However, when it comes to the ``softmaxing culture” problem, neither community seems to offer satisfying answers. HCI tends to suggest that ML does not adequately understand human-centered research; meanwhile, ML is quick to point out HCI’s technical gaps.} researchers across these fields are drawing on established social science criteria for evaluation \citep{hofstede_cultures_2001,hofstede_dimensionalizing_2011}, as well as constructing benchmarks informed by varied interpretations of language and culture \citep{shi_culturebank_2024,adelani_irokobench_2025,singh_global_2025,masoud_cultural_2024,ochieng_beyond_2024}. Indeed, across recent survey publications, the majority of papers examined in ML and NLP are drawing on methods such as Hofstede’s dimensions of culture and the World Values Survey. These methods have been instructive in generating useful and actionable insights into the workings of large AI models. While ML and HCI are adopting socio-technical approaches towards culturally aware models, there are limitations. These approaches are plagued by fundamental gaps in social and cultural understanding of the values and assumptions embedded in the data and benchmarks leveraged in the evaluation of large AI models. This position paper presents two conceptual shifts: first, the question “what is culture?” is the wrong question to start with in scoping and operationalizing \textit{culture} in LLMs; second, following \citet{wiredu_are_1995}'s claim that there are cultural universals, the more pragmatic question to ask is “when is culture?” The goal lies in understanding relations between universals and particulars, and when they become relationally valid and tenable. While there are growing demands emerging for context-driven and more situated evaluations beyond the artifact into real-world contexts, for the sake of this paper, I will focus on the conceptual aspects of culture and its formulation in evaluation. By re-orienting our understanding of culture, we will have a better understanding of assessing cultural values and constructs in large AI models at different stages of development and use. 

\section{Evaluating `culture': How ML and HCI have a lot in common than they acknowledge}

Both ML and HCI recognize that context matters, albeit differently. On one hand, ML research (since the deep learning revolution) has been preoccupied with advancing the frontiers of AI capabilities by building on the ``unreasonable effectiveness" of big data \citep{halevy_unreasonable_2009}. These advancements have been complemented by the rapid development and adoption of benchmarks to measure and compare modes across tasks and capabilities. Breakthroughs in ML, particularly modern approaches to embeddings,\footnote{Notably neural network-based approaches including \texttt{word2vec}, \texttt{seq2seq}, and \texttt{transformers} trained on enormous amounts of data.} showed meaningful connections between words and sentences, providing the capacity for modeling and measuring cultural contexts through large AI models. On the other hand, HCI has been engaged with understanding users in the contexts in which they interact with computing technologies. Sensitized in some part by \textit{the} generative model of culture, HCI examines how relationships between users and “technological objects and knowledge practices of everyday life become meaningful contingently and dynamically as social activity unfolds” \citep{irani_postcolonial_2010}.


But what is hardly surprising is that a growing body of recent work in ML and HCI shows limitations in the framing and operationalization of culture in system evaluations \citep{liu_culturally_2025,adilazuarda_towards_2024,zhou_culture_2025}. Whereas there is increasing recognition across the ML and HCI divide that culture is dynamic and relational, most evaluations still appear to capture narrow definitions and static snapshots of “culture.”


\section{What is culture? – and why it is the wrong question to begin with}

While there are several definitions of culture, they are out of scope of this paper. I want to focus on Hofstede's definition and taxonomy of culture that have achieved wide acceptance and acclaim in the social sciences since the 1970s and more recently in ML evaluation research.  According to \citet{hofstede_dimensionalizing_2011}, ``culture is the collective programming of the mind that distinguishes the members of one group or category of people from others." Culture is software for the society of minds. \citet{hofstede_cultures_2001,hofstede_dimensionalizing_2011} across several decades of work has developed a leading framework for evaluating culture: Hofstede's Dimensions of Culture \citep{hofstede_dimensionalizing_2011}. Hofstede’s cultural dimensions describe how societies -- countries and institutions -- vary in their values and practices, and these dimensions are categorized in six ways: power distance (acceptance of hierarchy); uncertainty avoidance (comfort with ambiguity); individualism versus collectivism (self vs. group); masculinity versus femininity (competition vs. care); and long-term versus short-term orientation (future vs.tradition), indulgence versus restraint (freedoms vs. control). 

NLP and sociolinguistics scholars \cite{zhou_culture_2025} have provided one of the most compelling cases against mainstream evaluations of culture in AI, suggesting that macro-level definitions of culture significantly constrain how culture is understood and evaluated. These definitions are unsatisfactory. Instead the scholars argue for “cultural competence” for NLP, with renewed attention to localization. Many grassroots efforts such as \textit{Masakhane} and \textit{Deep Learning Indaba} have made considerable progress and continue to show great promise on this path toward localizing and contextualizing language models \citep{nekoto_participatory_2020}. Indeed benchmarks and fine-tuning, often constructed from curations of data workers, Wikipedia articles, and corpora of community-specific social media posts, have advanced our understanding of cultural alignment in AI. However, culture is not about the identification of particular symbols, ideologies, cognitive patterns, etc. Culture is dynamic and complex. As such, attempting to nail down culture to a narrow definition -- captured in datasets -- only risks stereotyping and essentializing certain people (and their cultures) \citep{zhou_culture_2025,bender_dangers_2021}. The question “what is culture [in AI]?” is therefore the wrong question to start with. The more productive question to ask is ``when is culture?" This question challenges researchers to fully examine what it means for \textit{culture} to be relationally valid and tenable. Implicit in this question, too, is the partiality of our understanding of culture -- and how this must be accounted for in evaluations.

\section{Are there cultural universals?}

The short answer is \textit{yes}. For \citet{wiredu_are_1995,wiredu_cultural_1996}, a Ghanaian philosopher, if there were no cultural universals, there would be no intercultural communication. Cultural universals -- notably through language (an artifact of culture) -- enable the necessary translation work. Although his argument is simple, it is relevant to evaluations of culture in AI in at least two ways. First, humans generally share a `cognitive toolkit' that endows them with capabilities such as reflective perception (possession of a concept of the external world), abstraction (bringing particulars under general concepts), deduction (applying rules to observations), and induction (learning from examples).\footnote{While the deduction can be made that large AI models are also endowed a related `cognitive toolkit,' those arguments are beyond the scope of this paper.} Second, and importantly, humans possess the principle of sympathetic impartiality; a certain \textit{due} concern for others. Now our role here, as set up in the first point, is to identify the universals -- and of course the particulars. To be sure, this point is not about definitions, but rather demands that we think carefully about the human project; what makes us human, irrespective of our geographic, historical, moralistic, and even linguistic differences, and how we can account for these differences in our research. 

\section{Beyond `softmaxing’ culture: A more pragmatic research vision}
The use of the metaphor \textit{softmaxing culture} is closely related to the tendencies of LLMs to homogenize text and media. These tendencies are \textit{the} feature of transformer-based large AI systems, and they have exacerbated \textit{cultural incongruencies} of AI \citep{prabhakaran_cultural_2022}: the odds between the cultural assumptions baked into systems (both explicitly and from data sources) and the expectations of the target cultural ecosystems these technologies are used. Despite these structural issues, I argue that researchers in ML and HCI working around issues and questions of culture should focus on understanding the relational contexts through which cultural values and practices are made intelligible, especially for low-resource languages and expressions that are underrepresented in pre-training and fine-tuning datasets. Going \textit{local} is an early sign of success. However, confronting the \textit{softmaxing problem} will entail deep and meaningful integration of theoretical and empirical strategies for studying cultures beyond the ML/HCI divide.




\begin{thebibliography}{25}
\providecommand{\natexlab}[1]{#1}
\providecommand{\url}[1]{\texttt{#1}}
\expandafter\ifx\csname urlstyle\endcsname\relax
  \providecommand{\doi}[1]{doi: #1}\else
  \providecommand{\doi}{doi: \begingroup \urlstyle{rm}\Url}\fi

\bibitem[Adelani et~al.(2025)Adelani, Ojo, Azime, Zhuang, Alabi, He, Ochieng, Hooker, Bukula, Lee, Chukwuneke, Buzaaba, Sibanda, Kalipe, Mukiibi, Kabongo, Yuehgoh, Setaka, Ndolela, Odu, Mabuya, Muhammad, Osei, Samb, Guge, Sherman, and Stenetorp]{adelani_irokobench_2025}
David~Ifeoluwa Adelani, Jessica Ojo, Israel~Abebe Azime, Jian~Yun Zhuang, Jesujoba~O. Alabi, Xuanli He, Millicent Ochieng, Sara Hooker, Andiswa Bukula, En-Shiun~Annie Lee, Chiamaka Chukwuneke, Happy Buzaaba, Blessing Sibanda, Godson Kalipe, Jonathan Mukiibi, Salomon Kabongo, Foutse Yuehgoh, Mmasibidi Setaka, Lolwethu Ndolela, Nkiruka Odu, Rooweither Mabuya, Shamsuddeen~Hassan Muhammad, Salomey Osei, Sokhar Samb, Tadesse~Kebede Guge, Tombekai~Vangoni Sherman, and Pontus Stenetorp.
\newblock {IrokoBench}: {A} {New} {Benchmark} for {African} {Languages} in the {Age} of {Large} {Language} {Models}, January 2025.
\newblock URL \url{http://arxiv.org/abs/2406.03368}.
\newblock arXiv:2406.03368 [cs].

\bibitem[Adilazuarda et~al.(2024)Adilazuarda, Mukherjee, Lavania, Singh, Aji, O'Neill, Modi, and Choudhury]{adilazuarda_towards_2024}
Muhammad~Farid Adilazuarda, Sagnik Mukherjee, Pradhyumna Lavania, Siddhant Singh, Alham~Fikri Aji, Jacki O'Neill, Ashutosh Modi, and Monojit Choudhury.
\newblock Towards {Measuring} and {Modeling} "{Culture}" in {LLMs}: {A} {Survey}, September 2024.
\newblock URL \url{http://arxiv.org/abs/2403.15412}.
\newblock arXiv:2403.15412 [cs].

\bibitem[Agarwal et~al.(2025)Agarwal, Naaman, and Vashistha]{agarwal_ai_2025}
Dhruv Agarwal, Mor Naaman, and Aditya Vashistha.
\newblock {AI} {Suggestions} {Homogenize} {Writing} {Toward} {Western} {Styles} and {Diminish} {Cultural} {Nuances}.
\newblock In \emph{Proceedings of the 2025 {CHI} {Conference} on {Human} {Factors} in {Computing} {Systems}}, {CHI} '25, pages 1--21, New York, NY, USA, April 2025. Association for Computing Machinery.
\newblock ISBN 9798400713941.
\newblock \doi{10.1145/3706598.3713564}.
\newblock URL \url{https://dl.acm.org/doi/10.1145/3706598.3713564}.

\bibitem[Bajohr(2025)]{bajohr_thinking_2025}
Hannes Bajohr, editor.
\newblock \emph{Thinking with {AI}: {Machine} {Learning} the {Humanities}}.
\newblock Open Humanites Press, 2025.
\newblock ISBN 978-1-78542-141-9 978-1-78542-140-2.
\newblock URL \url{https://www.openhumanitiespress.org/books/titles/thinking-with-ai/}.

\bibitem[Bender et~al.(2021)Bender, Gebru, McMillan-Major, and Shmitchell]{bender_dangers_2021}
Emily~M. Bender, Timnit Gebru, Angelina McMillan-Major, and Shmargaret Shmitchell.
\newblock On the {Dangers} of {Stochastic} {Parrots}: {Can} {Language} {Models} {Be} {Too} {Big}?
\newblock In \emph{Proceedings of the 2021 {ACM} {Conference} on {Fairness}, {Accountability}, and {Transparency}}, {FAccT} '21, pages 610--623, New York, NY, USA, March 2021. Association for Computing Machinery.
\newblock ISBN 978-1-4503-8309-7.
\newblock \doi{10.1145/3442188.3445922}.
\newblock URL \url{https://doi.org/10.1145/3442188.3445922}.

\bibitem[Bommasani et~al.(2022)Bommasani, Hudson, Adeli, Altman, Arora, Arx, Bernstein, Bohg, Bosselut, Brunskill, Brynjolfsson, Buch, Card, Castellon, Chatterji, Chen, Creel, Davis, Demszky, Donahue, Doumbouya, Durmus, Ermon, Etchemendy, Ethayarajh, Fei-Fei, Finn, Gale, Gillespie, Goel, Goodman, Grossman, Guha, Hashimoto, Henderson, Hewitt, Ho, Hong, Hsu, Huang, Icard, Jain, Jurafsky, Kalluri, Karamcheti, Keeling, Khani, Khattab, Koh, Krass, Krishna, Kuditipudi, Kumar, Ladhak, Lee, Lee, Leskovec, Levent, Li, Li, Ma, Malik, Manning, Mirchandani, Mitchell, Munyikwa, Nair, Narayan, Narayanan, Newman, Nie, Niebles, Nilforoshan, Nyarko, Ogut, Orr, Papadimitriou, Park, Piech, Portelance, Potts, Raghunathan, Reich, Ren, Rong, Roohani, Ruiz, Ryan, Ré, Sadigh, Sagawa, Santhanam, Shih, Srinivasan, Tamkin, Taori, Thomas, Tramèr, Wang, Wang, Wu, Wu, Wu, Xie, Yasunaga, You, Zaharia, Zhang, Zhang, Zhang, Zhang, Zheng, Zhou, and Liang]{bommasani_opportunities_2022}
Rishi Bommasani, Drew~A. Hudson, Ehsan Adeli, Russ Altman, Simran Arora, Sydney~von Arx, Michael~S. Bernstein, Jeannette Bohg, Antoine Bosselut, Emma Brunskill, Erik Brynjolfsson, Shyamal Buch, Dallas Card, Rodrigo Castellon, Niladri Chatterji, Annie Chen, Kathleen Creel, Jared~Quincy Davis, Dora Demszky, Chris Donahue, Moussa Doumbouya, Esin Durmus, Stefano Ermon, John Etchemendy, Kawin Ethayarajh, Li~Fei-Fei, Chelsea Finn, Trevor Gale, Lauren Gillespie, Karan Goel, Noah Goodman, Shelby Grossman, Neel Guha, Tatsunori Hashimoto, Peter Henderson, John Hewitt, Daniel~E. Ho, Jenny Hong, Kyle Hsu, Jing Huang, Thomas Icard, Saahil Jain, Dan Jurafsky, Pratyusha Kalluri, Siddharth Karamcheti, Geoff Keeling, Fereshte Khani, Omar Khattab, Pang~Wei Koh, Mark Krass, Ranjay Krishna, Rohith Kuditipudi, Ananya Kumar, Faisal Ladhak, Mina Lee, Tony Lee, Jure Leskovec, Isabelle Levent, Xiang~Lisa Li, Xuechen Li, Tengyu Ma, Ali Malik, Christopher~D. Manning, Suvir Mirchandani, Eric Mitchell, Zanele Munyikwa, Suraj Nair,
  Avanika Narayan, Deepak Narayanan, Ben Newman, Allen Nie, Juan~Carlos Niebles, Hamed Nilforoshan, Julian Nyarko, Giray Ogut, Laurel Orr, Isabel Papadimitriou, Joon~Sung Park, Chris Piech, Eva Portelance, Christopher Potts, Aditi Raghunathan, Rob Reich, Hongyu Ren, Frieda Rong, Yusuf Roohani, Camilo Ruiz, Jack Ryan, Christopher Ré, Dorsa Sadigh, Shiori Sagawa, Keshav Santhanam, Andy Shih, Krishnan Srinivasan, Alex Tamkin, Rohan Taori, Armin~W. Thomas, Florian Tramèr, Rose~E. Wang, William Wang, Bohan Wu, Jiajun Wu, Yuhuai Wu, Sang~Michael Xie, Michihiro Yasunaga, Jiaxuan You, Matei Zaharia, Michael Zhang, Tianyi Zhang, Xikun Zhang, Yuhui Zhang, Lucia Zheng, Kaitlyn Zhou, and Percy Liang.
\newblock On the {Opportunities} and {Risks} of {Foundation} {Models}, July 2022.
\newblock URL \url{http://arxiv.org/abs/2108.07258}.
\newblock arXiv:2108.07258 [cs].

\bibitem[Cheng et~al.(2023)Cheng, Durmus, and Jurafsky]{cheng_marked_2023}
Myra Cheng, Esin Durmus, and Dan Jurafsky.
\newblock Marked {Personas}: {Using} {Natural} {Language} {Prompts} to {Measure} {Stereotypes} in {Language} {Models}, May 2023.
\newblock URL \url{http://arxiv.org/abs/2305.18189}.
\newblock arXiv:2305.18189 [cs].

\bibitem[Decoupes et~al.(2025)Decoupes, Interdonato, Roche, Teisseire, and Valentin]{decoupes_evaluation_2025}
Rémy Decoupes, Roberto Interdonato, Mathieu Roche, Maguelonne Teisseire, and Sarah Valentin.
\newblock Evaluation of {Geographical} {Distortions} in {Language} {Models}: {A} {Crucial} {Step} {Towards} {Equitable} {Representations}.
\newblock volume 15243, pages 86--100. 2025.
\newblock \doi{10.1007/978-3-031-78977-9_6}.
\newblock URL \url{http://arxiv.org/abs/2404.17401}.
\newblock arXiv:2404.17401 [cs].

\bibitem[Farrell et~al.(2025)Farrell, Gopnik, Shalizi, and Evans]{farrell_large_2025}
Henry Farrell, Alison Gopnik, Cosma Shalizi, and James Evans.
\newblock Large {AI} models are cultural and social technologies.
\newblock \emph{Science}, 387\penalty0 (6739):\penalty0 1153--1156, March 2025.
\newblock \doi{10.1126/science.adt9819}.
\newblock URL \url{https://www.science.org/doi/10.1126/science.adt9819}.
\newblock Publisher: American Association for the Advancement of Science.

\bibitem[Halevy et~al.(2009)Halevy, Norvig, and Pereira]{halevy_unreasonable_2009}
Alon Halevy, Peter Norvig, and Fernando Pereira.
\newblock The {Unreasonable} {Effectiveness} of {Data}.
\newblock \emph{IEEE Intelligent Systems}, 24\penalty0 (2):\penalty0 8--12, March 2009.
\newblock ISSN 1541-1672.
\newblock \doi{10.1109/MIS.2009.36}.
\newblock URL \url{http://ieeexplore.ieee.org/document/4804817/}.

\bibitem[Hofstede(2001)]{hofstede_cultures_2001}
Geert Hofstede.
\newblock \emph{Culture's {Consequences}: {Comparing} {Values}, {Behaviors}, {Institutions} and {Organizations} {Across} {Nations}}.
\newblock SAGE, second edition, 2001.
\newblock ISBN 978-0-8039-7324-4.
\newblock Google-Books-ID: w6z18LJ\_1VsC.

\bibitem[Hofstede(2011)]{hofstede_dimensionalizing_2011}
Geert Hofstede.
\newblock Dimensionalizing {Cultures}: {The} {Hofstede} {Model} in {Context}.
\newblock \emph{Online Readings in Psychology and Culture}, 2\penalty0 (1), December 2011.
\newblock ISSN 2307-0919.
\newblock \doi{10.9707/2307-0919.1014}.
\newblock URL \url{https://scholarworks.gvsu.edu/orpc/vol2/iss1/8}.

\bibitem[Irani et~al.(2010)Irani, Vertesi, Dourish, Philip, and Grinter]{irani_postcolonial_2010}
Lilly Irani, Janet Vertesi, Paul Dourish, Kavita Philip, and Rebecca~E. Grinter.
\newblock Postcolonial computing: a lens on design and development.
\newblock In \emph{Proceedings of the {SIGCHI} {Conference} on {Human} {Factors} in {Computing} {Systems}}, {CHI} '10, pages 1311--1320, New York, NY, USA, April 2010. Association for Computing Machinery.
\newblock ISBN 978-1-60558-929-9.
\newblock \doi{10.1145/1753326.1753522}.
\newblock URL \url{https://doi.org/10.1145/1753326.1753522}.

\bibitem[Liu et~al.(2025)Liu, Gurevych, and Korhonen]{liu_culturally_2025}
Chen~Cecilia Liu, Iryna Gurevych, and Anna Korhonen.
\newblock Culturally {Aware} and {Adapted} {NLP}: {A} {Taxonomy} and a {Survey} of the {State} of the {Art}, March 2025.
\newblock URL \url{http://arxiv.org/abs/2406.03930}.
\newblock arXiv:2406.03930 [cs].

\bibitem[Masoud et~al.(2024)Masoud, Liu, Ferianc, Treleaven, and Rodrigues]{masoud_cultural_2024}
Reem~I. Masoud, Ziquan Liu, Martin Ferianc, Philip Treleaven, and Miguel Rodrigues.
\newblock Cultural {Alignment} in {Large} {Language} {Models}: {An} {Explanatory} {Analysis} {Based} on {Hofstede}'s {Cultural} {Dimensions}, May 2024.
\newblock URL \url{http://arxiv.org/abs/2309.12342}.
\newblock arXiv:2309.12342 [cs].

\bibitem[Nekoto et~al.(2020)Nekoto, Marivate, Matsila, Fasubaa, Fagbohungbe, Akinola, Muhammad, Kabongo~Kabenamualu, Osei, Sackey, Niyongabo, Macharm, Ogayo, Ahia, Berhe, Adeyemi, Mokgesi-Selinga, Okegbemi, Martinus, Tajudeen, Degila, Ogueji, Siminyu, Kreutzer, Webster, Ali, Abbott, Orife, Ezeani, Dangana, Kamper, Elsahar, Duru, Kioko, Espoir, Van~Biljon, Whitenack, Onyefuluchi, Emezue, Dossou, Sibanda, Bassey, Olabiyi, Ramkilowan, Öktem, Akinfaderin, and Bashir]{nekoto_participatory_2020}
Wilhelmina Nekoto, Vukosi Marivate, Tshinondiwa Matsila, Timi Fasubaa, Taiwo Fagbohungbe, Solomon~Oluwole Akinola, Shamsuddeen Muhammad, Salomon Kabongo~Kabenamualu, Salomey Osei, Freshia Sackey, Rubungo~Andre Niyongabo, Ricky Macharm, Perez Ogayo, Orevaoghene Ahia, Musie~Meressa Berhe, Mofetoluwa Adeyemi, Masabata Mokgesi-Selinga, Lawrence Okegbemi, Laura Martinus, Kolawole Tajudeen, Kevin Degila, Kelechi Ogueji, Kathleen Siminyu, Julia Kreutzer, Jason Webster, Jamiil~Toure Ali, Jade Abbott, Iroro Orife, Ignatius Ezeani, Idris~Abdulkadir Dangana, Herman Kamper, Hady Elsahar, Goodness Duru, Ghollah Kioko, Murhabazi Espoir, Elan Van~Biljon, Daniel Whitenack, Christopher Onyefuluchi, Chris~Chinenye Emezue, Bonaventure F.~P. Dossou, Blessing Sibanda, Blessing Bassey, Ayodele Olabiyi, Arshath Ramkilowan, Alp Öktem, Adewale Akinfaderin, and Abdallah Bashir.
\newblock Participatory {Research} for {Low}-resourced {Machine} {Translation}: {A} {Case} {Study} in {African} {Languages}.
\newblock In \emph{Findings of the {Association} for {Computational} {Linguistics}: {EMNLP} 2020}, pages 2144--2160, Online, 2020. Association for Computational Linguistics.
\newblock \doi{10.18653/v1/2020.findings-emnlp.195}.
\newblock URL \url{https://www.aclweb.org/anthology/2020.findings-emnlp.195}.

\bibitem[Ochieng et~al.(2024)Ochieng, Gumma, Sitaram, Wang, Chaudhary, Ronen, Bali, and O'Neill]{ochieng_beyond_2024}
Millicent Ochieng, Varun Gumma, Sunayana Sitaram, Jindong Wang, Vishrav Chaudhary, Keshet Ronen, Kalika Bali, and Jacki O'Neill.
\newblock Beyond {Metrics}: {Evaluating} {LLMs}' {Effectiveness} in {Culturally} {Nuanced}, {Low}-{Resource} {Real}-{World} {Scenarios}, June 2024.
\newblock URL \url{http://arxiv.org/abs/2406.00343}.
\newblock arXiv:2406.00343 [cs].

\bibitem[Prabhakaran et~al.(2022)Prabhakaran, Qadri, and Hutchinson]{prabhakaran_cultural_2022}
Vinodkumar Prabhakaran, Rida Qadri, and Ben Hutchinson.
\newblock Cultural {Incongruencies} in {Artificial} {Intelligence}, November 2022.
\newblock URL \url{http://arxiv.org/abs/2211.13069}.
\newblock arXiv:2211.13069 [cs].

\bibitem[Shi et~al.(2024)Shi, Li, Zhang, Ziems, yu, Horesh, Paula, and Yang]{shi_culturebank_2024}
Weiyan Shi, Ryan Li, Yutong Zhang, Caleb Ziems, Chunhua yu, Raya Horesh, Rogério Abreu~de Paula, and Diyi Yang.
\newblock {CultureBank}: {An} {Online} {Community}-{Driven} {Knowledge} {Base} {Towards} {Culturally} {Aware} {Language} {Technologies}, April 2024.
\newblock URL \url{http://arxiv.org/abs/2404.15238}.
\newblock arXiv:2404.15238 [cs].

\bibitem[Singh et~al.(2025)Singh, Romanou, Fourrier, Adelani, Ngui, Vila-Suero, Limkonchotiwat, Marchisio, Leong, Susanto, Ng, Longpre, Ko, Ruder, Smith, Bosselut, Oh, Martins, Choshen, Ippolito, Ferrante, Fadaee, Ermis, and Hooker]{singh_global_2025}
Shivalika Singh, Angelika Romanou, Clémentine Fourrier, David~I. Adelani, Jian~Gang Ngui, Daniel Vila-Suero, Peerat Limkonchotiwat, Kelly Marchisio, Wei~Qi Leong, Yosephine Susanto, Raymond Ng, Shayne Longpre, Wei-Yin Ko, Sebastian Ruder, Madeline Smith, Antoine Bosselut, Alice Oh, Andre F.~T. Martins, Leshem Choshen, Daphne Ippolito, Enzo Ferrante, Marzieh Fadaee, Beyza Ermis, and Sara Hooker.
\newblock Global {MMLU}: {Understanding} and {Addressing} {Cultural} and {Linguistic} {Biases} in {Multilingual} {Evaluation}, February 2025.
\newblock URL \url{http://arxiv.org/abs/2412.03304}.
\newblock arXiv:2412.03304 [cs].

\bibitem[Wallach et~al.(2025)Wallach, Desai, Cooper, Wang, Atalla, Barocas, Blodgett, Chouldechova, Corvi, Dow, Garcia-Gathright, Olteanu, Pangakis, Reed, Sheng, Vann, Vaughan, Vogel, Washington, and Jacobs]{wallach_position_2025}
Hanna Wallach, Meera Desai, A.~Feder Cooper, Angelina Wang, Chad Atalla, Solon Barocas, Su~Lin Blodgett, Alexandra Chouldechova, Emily Corvi, P.~Alex Dow, Jean Garcia-Gathright, Alexandra Olteanu, Nicholas Pangakis, Stefanie Reed, Emily Sheng, Dan Vann, Jennifer~Wortman Vaughan, Matthew Vogel, Hannah Washington, and Abigail~Z. Jacobs.
\newblock Position: {Evaluating} {Generative} {AI} {Systems} is a {Social} {Science} {Measurement} {Challenge}, February 2025.
\newblock URL \url{http://arxiv.org/abs/2502.00561}.
\newblock arXiv:2502.00561 [cs].

\bibitem[Wiredu(1995)]{wiredu_are_1995}
Kwasi Wiredu.
\newblock Are {There} {Cultural} {Universals}?
\newblock \emph{The Monist}, 78\penalty0 (1):\penalty0 52--64, 1995.
\newblock ISSN 0026-9662.
\newblock URL \url{https://www.jstor.org/stable/27903418}.
\newblock Publisher: Oxford University Press.

\bibitem[Wiredu(1996)]{wiredu_cultural_1996}
Kwasi Wiredu.
\newblock \emph{Cultural universals and particulars: {An} {African} perspective}.
\newblock African {Systems} of {Thought}. Indiana University Press, Bloomington, January 1996.
\newblock ISBN 978-0-253-21080-7.
\newblock URL \url{https://research.ebsco.com/linkprocessor/plink?id=b733d0e7-ab78-3554-b888-33529bb83cf5}.

\bibitem[Ye et~al.(2024)Ye, Wang, Huang, Chen, Zhang, Moniz, Gao, Geyer, Huang, Chen, Chawla, and Zhang]{ye_justice_2024}
Jiayi Ye, Yanbo Wang, Yue Huang, Dongping Chen, Qihui Zhang, Nuno Moniz, Tian Gao, Werner Geyer, Chao Huang, Pin-Yu Chen, Nitesh~V. Chawla, and Xiangliang Zhang.
\newblock Justice or {Prejudice}? {Quantifying} {Biases} in {LLM}-as-a-{Judge}, October 2024.
\newblock URL \url{http://arxiv.org/abs/2410.02736}.
\newblock arXiv:2410.02736 [cs].

\bibitem[Zhou et~al.(2025)Zhou, Bamman, and Bleaman]{zhou_culture_2025}
Naitian Zhou, David Bamman, and Isaac~L. Bleaman.
\newblock Culture is {Not} {Trivia}: {Sociocultural} {Theory} for {Cultural} {NLP}, February 2025.
\newblock URL \url{http://arxiv.org/abs/2502.12057}.
\newblock arXiv:2502.12057 [cs].

\end{thebibliography}

\end{document}